\documentstyle[epsf,rotate,twocolumn,aps]{revtex}

\begin{document}
\draft              

\twocolumn[\hsize\textwidth\columnwidth\hsize\csname 
@twocolumnfalse\endcsname


\title{
A self-consistent, conserving theory of the\\
attractive Hubbard model in two dimensions}

\author{M. Letz$^*$ and R. J. Gooding}

\address{Dept. of Physics, Queen's University,\\ Kingston, ON Canada K7L 3N6}
\address{$^*$ present address: 
Institut f\"ur Physik,\\
Johannes-Gutenberg Universit\"at,
55099 Mainz, Germany}
\date{\today}

\maketitle

\begin{abstract}

We have investigated the attractive Hubbard model in the low density
limit for the 2D square lattice using the ladder approximation
for the vertex function in a self-consistent, conserving formulation.
In the parameter region where the on-site attraction is of the order
of the bandwidth, we found no evidence of a pseudo gap. Further,
we have observed that the suppression of the Fermi surface known
to destroy superconductivity in one and two dimensions, when
these systems are treated using a non self-consistent theory 
(Schmitt-Rink, {\it et al.}, Phys. Rev. Lett. {\bf 63}, 445 (1989)), 
does not occur when pair-pair interactions are included.
However, we do find a quasiparticle lifetime that varies linearly with 
temperature, similar to many experiments. Thus, although this system
has a Fermi surface, it shows non Fermi liquid type behaviour 
over a wide temperature range. We stress that our work
uses thermal Green's functions along the real time axis, and thus allows 
for a more accurate determination of the dynamical properties of a
model than theories that require extrapolations from the imaginary frequency 
axis.

\end{abstract}
]
\vskip -0.5cm
\narrowtext

\section{Introduction:}
\label{sec:intro}

The high temperature superconductors show remarkable deviations from
Fermi-liquid behaviour in their normal state above T$_c$,
the superconducting transition temperature. Although
this seems to be experimentally well established, no consistent
microscopic theoretical explanation has been found. 

There are a number of major differences with usual metallic (BCS-type) 
superconductors on which we concentrate for a theoretical understanding. 
The first one is the low dimensionality. For example, the normal state 
conductivity mainly takes place in the two-dimensional ($d$-dimensions 
shall be denoted by $d$D throughout this paper) copper oxide planes.
The second difference is the extremely short coherence
length of the Cooper pairs in the superconducting state. These are
known to be of the order of 20 $\AA$ (3-4 lattice constants), and 
therefore much smaller than in usual superconductors ($\sim 1000 \AA$). 
This fact, together with the extremely low (''bad metals'') quasiparticle 
density leads to pairs which are barely overlapping ($\sim 10^{-2}
$ pairs / coherence volume) with each other. Such arguments were first
stated in detail by Randeria \cite{randeriarev} who investigated
conditions for a crossover between superconductivity and Bose
condensation of pairs of electrons. The small overlap is claimed to
be related to a separation of a temperature T$^*$, at which pairing
takes place, from the temperature T$_c$, at which phase coherence and
therefore superconductivity is established. In contrast to this, usual
superconductors have $\sim 10^6$ pairs / coherence volume which is
believed to be the reason why pairing and phase coherence take place
at the same temperature, the mean-field T$_c$. Another unusual
property of the cuprate superconductors is the linear
resistivity in the normal state of the optimally doped materials. An
ideal Fermi liquid should show a resistivity caused by
electron-electron scattering near the Fermi surface which varies
quadratically with the temperature. On the other hand, a crossover to a
linear resistivity due to phonons is expected to take place at much
higher temperatures, above the Debye temperature.

Finally, one last observation that we wish to focus on is the presence
of a pseudo gap. This feature was first
observed with NMR \cite{warren89,takigawa91,alloul93} by measuring the
spin-lattice relaxation rate for the $^{65}$Cu nucleus. The relaxation rate,
$\frac{1}{T_1\;T}$, which is temperature independent in normal metals,
decreases strongly with decreasing T even above T$_c$, in
the high-T$_c$-cuprates. The pseudogap has also been measured by optical
experiments where it is found in the temperature dependence of the
scattering time obtained from a generalized Drude theory for optical
conductivity data \cite{pseudo,timusk96}. The momentum dependence of the pseudo
gap has been investigated with recent ARPES experiments, and it is
found to be present mainly along the $(k,0)$-direction, consistent
with the proposed $d$-wave symmetry of the superconducting order parameter. 
It has also been observed by $\mu$sR \cite{musR_pseu} and STM \cite{stm_pseu}
measurements. At present, there is ongoing discussion addressing the
notion that the gap can be related to pair formation, or other possible 
precursor phenomena of superconductivity, which takes place at temperatures 
above that at which macroscopic phase coherence is established. 

The above-mentioned ideas have led us to consider a simple model system
in 2D which can describe short coherence length pairs which might exist
as preformed pairs above T$_c$. This model, the attractive Hubbard model,
for on-site, $s$-wave pairing only, allows us to focus on many of
the above-mentioned properties. In particular, we have examined the
dynamical properties of this model to see if the attractive Hubbard
model in 2D possesses a pseudogap. Further, we have examined the 
temperature dependence of the imaginary part of the single-particle 
self energy to learn how the scattering rate behaves.

We have focussed on this model in the low density limit. When
studied in the limit of low band filling the attractive Hubbard model 
represents a system with low quasiparticle densities, and therefore the 
weakly overlapping pairs proposed to characterize the cuprate 
superconductors are a natural consequence of this problem.
An approximation which works well in the dilute limit
is the ladder approximation; this formalism accounts for all possible
scattering events between particles that can occur in the particle-particle
channel (only particle-hole scatterings are ignored). 

In a simple and elegant paper \cite {svr}, Schmitt-Rink {\it et al.}, studied 
such model systems and concluded, at least in a non self-consistent treatment
of such problems, that in 2D a stable, 2-particle bound state persists 
down to T=0, and this leads to a T=0 Bose condensation of composite
bosons (two fermions pair to form a boson). The physics of this
phenomenon is that the appearance of preformed pairs leads to
the elimination of the Fermi surface (there are no fermions left),
and therefore superconductivity is suppressed in favour of T=0 Bose 
condensation.

The consequence of this work to the model system under consideration
is as follows: in a non self-consistent treatment of the 
attractive Hubbard model at low densities, employing the ladder 
approximation, the system is always unstable towards T=0 Bose condensation 
of preformed pairs into an infinite lifetime, two-particle 
bound state \cite {svr}. One of the focuses of this paper, and a second 
motivating force behind our study, is to test if this idea survives when 
the theory is solved in a fully self-consistent fashion. That is, when one
includes interactions between pairs, does the physics of Ref. \cite {svr} 
survive? A careful and detailed study of the interaction between
pairs was given by Haussmann \cite {haussmann93,haussmann94}, and here we will
consider these ideas as applied to the observed normal state anomalies,
including the formation of a pseudogap, and to the physics of Ref.
\cite {svr}.

The numerical work of Haussmann in 3D suggests that the physics of
this problem is very different than that of its non self-consistent
counterpart. In particular, he suggests that the bound state
strongly hybridizes with the two-particle scattering continuum, and
that this greatly reduces the appearance of preformed pairs.
Such a tendency has also been suggested by the work of several other
authors. Fresard {\it et al.} \cite{fresard92} found in 2D, by
applying the selfconsistent ladder approximation, that in the low doping 
regime the Fermi-liquid properties are fully recovered and that
only in the strong coupling regime can deviations from Fermi-liquid
behaviour be expected. (The work of these authors differs from our
own work in that they use a very different method to obtain the
dynamical properties of this model in this approximation.) Also, 
Singer {\it et al.}, \cite{singer96} employed
quantum Monte Carlo methods and concluded that the two-particle bound
state, which they refer to as a ``band of pairs'', is strongly overlapping
with the one particle continuum, and only at very large attractive
interactions does it become well separated from the one particle continuum.
Lastly, recent analytical work of Kagan {\it et al.} \cite{kagan97} 
shows that when pair-pair interactions are included in a T=0 calculation, 
in the dilute limit the gap between the two-particle bound state and the 
one-particle continuum starts to close.

Our results are consistent with the above-state trends, and
we do not find that the physics of Ref. \cite {svr} survives
when pair-pair interactions are included. Further,
we find no evidence of a pseudogap. We have used double-time
Green's functions to study this system at T$>$0, and thus unlike
other studies that examined this system using imaginary times
(the conventional Matsubara frequency formulation) who had
to rely on Pad\'e approximants or maximum entropy techniques,
we are able to examine the dynamical properties of this
system directly. Thus, we believe that we have a somewhat
more reliable representation of the dynamics of this system.

We organize our paper as follows. In \S \ref{sec:model} we
introduce the model and describe different levels of approximation
which can be used for the ladder approximation. 
In the next section we introduce a ${\bf k}$-averaged method along
with a generalized spectral representation of all temperature-dependent
Green's functions which enables
us to obtain results in a fully self-consistent calculation along
the real time axis. In \S \ref{subsec:justify} we present numerical
evidence that aids in justifying these approximations.
Our main results are presented in \S \ref{sec:results}, and in \S
\ref{sec:conclusion} we present our conclusions.

\par\vfill\eject
\section{Model}
\label{sec:model}

The model Hamiltonian which we consider is the attractive (negative-U) 
Hubbard model, given by
\begin{equation}
H =-t\sum_{\langle ij \rangle, \sigma} 
( c^\dagger_{i,\sigma} c_{j,\sigma} + {\rm {h.c.}} )
~-~\mid U \mid~\sum_i n_{i,\uparrow} n_{i,\downarrow}~~,
\label{eq:NUHM}
\end{equation}
where $ t $ is the transfer integral between neighbouring lattice
sites $i,j$, $c^\dagger_{i,\sigma}, c_{i,\sigma}$ are electronic
creation, annihilation operators, respectively, and $\mid U \mid$ 
is the strength of the  on-site attractive interaction between two 
electrons occupying the same lattice site. Throughout this paper
we restrict our attention to $d$-dimensional hyper-cubic lattices.

For completeness, we begin by reviewing the well studied ladder
approximation to the Bethe-Salpeter equation \cite{fetwal131ff,haussmann93}.
The Dyson equation which has to be solved to obtain the full
one-particle Green's function contains a large number of complicated
diagrams and cannot be solved exactly. However, in the ladder 
approximation, which accounts for all possible
scattering events between particles that can occur in the particle-particle
channel (only particle-hole scatterings are ignored), one can find 
a solution for the single-particle self energy. The ladder approximation 
can be motivated in the dilute limit by taking $k_F~a$, 
the Fermi momentum multiplied with the scattering length (which in 3D
is given by $a = \frac{m \mid U \mid L^3}{4 \pi \hbar^2}$, 
which is the effective range of the attractive potential)  
as an additional small parameter. This is so because all diagrams which 
include more than one hole propagator (crossing diagrams) are neglected,
and therefore this approximation is valid in the low density
limit. The repeated scattering enters the equation through the vertex
function $\Gamma(K,i\Omega_n)$, and since the interaction $\mid U \mid$
is constant, the Bethe-Salpeter equation which determines 
$\Gamma(K,i\Omega_n)$ becomes exactly solvable.

\begin{figure}
\unitlength1cm
\epsfxsize=9cm
\begin{picture}(7,4.5)
\put(-1.0,-5.5){\epsffile{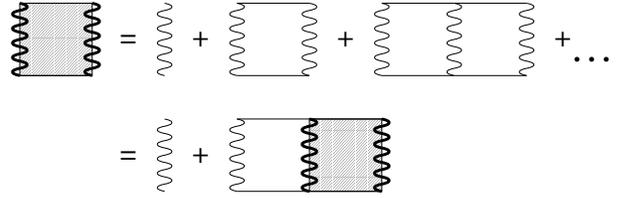}}
\end{picture}
\caption{Diagrammatic representation of Eq. (\protect\ref{eq:gammatild}). The
vertex function $\Gamma$ contains the repeated scattering of two particles.}
\label{fig:ladder}
\end{figure}

To display this solution we introduce the pair susceptibility, 
$\chi({\bf K},i \Omega _n)$, given by
\begin{equation}
\label{eq:chi}
\chi({\bf K},i \Omega _n) = -\frac{1}{N \beta} \sum_{m,{\bf k}}  
G({\bf K}-{\bf k},i \Omega _n - i \omega _m) G({\bf k},i \omega _m) 
\end{equation}
where $G({\bf k},i \omega _m)$ is the one-particle thermal Green's function. 
Note that the sign of this function is chosen to be different
by different authors, and readers should take note the consequence
of this sign choice in future equations.
Then, we can express the solution for the vertex function as 
\begin{equation}
\label{eq:gammatild}
\tilde{\Gamma} ({\bf K},i \Omega _n) =  - \mid U \mid  / 
\left ( 1 + \mid U \mid  \; \chi({\bf K},i \Omega _n) \right )~~,
\end{equation}
which is shown diagrammatically in Fig. \ref{fig:ladder}.

In our work we have chosen to subtract the interaction strength
from $\tilde{\Gamma}$ to obtain an analytic function $\Gamma({\bf K},z)$
with properties appropriate for examination by a generalized 
Kramers-Kroning analysis (this is equivalent to subtracting
the Hartree potential from the self energy), and thus introduce
\begin{eqnarray}
\label{eq:gamma}
\Gamma({\bf K},i \Omega _n) &=& \tilde{\Gamma}({\bf K},i \Omega _n) - 
(- \mid  U \mid) \nonumber \\
&=& \frac{U^2 \chi({\bf K},i \Omega _n)  }{ 
 1 + \mid U \mid  \; \chi({\bf K},i \Omega _n) }~~.  
\end{eqnarray}
From now on, we shall only refer to $\Gamma$.

The criterion which determines a breakdown of the normal state 
due to superconducting pair formation with decreasing temperature is
known as the Thouless criterion \cite{thoulesscrit}.  We have
examined this condition in detail, and will report on our
results elsewhere. Here, we simply remark
that the Thouless condition is associated with the occurrence of a
two-particle bound state (with infinite lifetime) at the chemical
potential, and is signified by
\begin{equation}
\label{eq:thlcrit}
1 + \mid U \mid  \; \chi({\bf K = 0}, z = 0) = 0~~.
\end{equation}

This equation, and the associated normal state properties, can
be examined in a variety of increasingly more accurate 
approximations, and we review these approximations before proceeding
to our results.

\subsection{Non self-consistent, non-conserving theory}
\label{subsec:11}

The self energy in this case (from now on denoted by NSCNC) is given by
\begin{equation}
\label{eq:selfnoncons}
\Sigma^0({\bf k},i \omega _n) =  \frac{1}{N \beta} \sum_{m,{\bf q}}
\Gamma^0 ({\bf k}+{\bf q },i \omega _m + 
i \omega _n) G^0({\bf q},i \omega _m)~~.
\end{equation}
The index $^0$ indicates the use of free Green's functions.
The full Green's function in this approximation is:
\begin{eqnarray}
\label{eq:greennoncons}
G({\bf k},i \omega _n) &=& G^0({\bf k},i \omega _n) 
\nonumber \\ &&+ G^0({\bf k},i
\omega _n) \Sigma^0({\bf k},i
\omega _n) G^0({\bf k},i \omega _n)~~,
\end{eqnarray}

Most importantly, Schmitt-Rink {\it et al.}, \cite{svr} have used this level 
of approximation and have shown that for any 2D system with an attractive 
interaction the system is unstable to the effective emptying of the Fermi 
''circle'' into the two-particle bound state. Then, a condensation at T=0 of 
these noninteracting composite bosons takes place. While some criticism
of this simple and elegant idea exists \cite {randeriarev},
we believe that for any non self-consistent theory the ideas of Ref. \cite
{svr} are solid. What happens when one includes
self consistency, namely when one includes pair interactions, is
one of the focuses of this paper. Further, unlike Ref. \cite {randeriarev},
we believe that it is more appropriate to assess the credibility of the 
ideas of Ref. \cite {svr} using the same kind of formalism.

\subsection{Non self-consistent, conserving theory}
\label{subsec:12}

\begin{figure}
\unitlength1cm
\epsfxsize=9cm
\begin{picture}(7,4.5)
\put(-0.5,-5.5){\epsffile{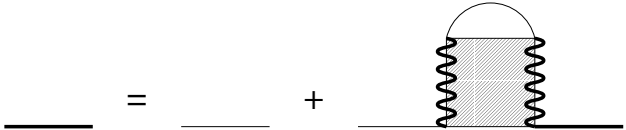}}
\end{picture}
\caption{Diagram for the single-particle Green's function (solid line) in the 
non self-consistent, conserving approximation (NSCC). The thin solid lines
represent the non-interacting Green's function.}
\label{fig:G4NSCC}
\end{figure}

We are studying a lattice model which has particle-hole symmetry.
Unfortunately, the NSCNC approximation violates this
symmetry. It can be restored if instead of Eq.~(\ref{eq:greennoncons})
for the Green's function we use
\begin{equation}
\label{eq:greencons}
G({\bf k},i \omega _n) = \left ( G^0({\bf k},i \omega _n)^{-1} - 
\Sigma^0({\bf k},i
\omega _n) \right ) ^{-1}
\end{equation}
The diagram for the full Green's function is shown in Fig. \ref{fig:G4NSCC}
and from now on we refer to this level of approximation as NSCC.

It was suggested by Serene \cite {serene89} that the inclusion
of these new diagrams fundamentally changes the physics of Ref. \cite {svr}.
At least for the attractive Hubbard model, we do not agree with
this claim, and instead find that the physics of Ref. \cite {svr}
is still correct and, in fact, is greatly simplified when one
uses a NSCC.

\subsection{Self-consistent, conserving theory}
\label{subsec:13}

\begin{figure}
\unitlength1cm
\epsfxsize=9cm
\begin{picture}(7,4.5)
\put(-0.5,-5.5){\epsffile{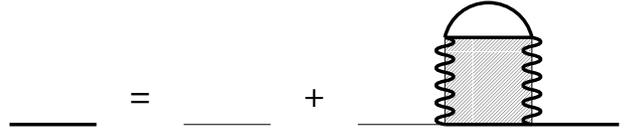}}
\end{picture}
\caption{Diagram for the single-particle Green's function in the 
self-consistent, conserving approximation (SCC), with the same notation
as Figs. 1 and 2.}
\label{fig:G4SCC}
\end{figure}

One may treat the ladder approximation fully self consistently by
calculating the self energy using the full interacting Green's
function. That is, one solves for
\begin{eqnarray}
\Sigma ({\bf k},i \omega _n) &=&  \frac{1}{N \beta} \sum_{m,{\bf q}}
\Gamma ({\bf k}+{\bf q },i \omega _m + 
i \omega _n) G({\bf q},i \omega _m) \label{eq:sigscf} \\
G({\bf k},i \omega _n) &=& \left ( G^0({\bf k},i \omega _n)^{-1} - 
\Sigma({\bf k},i
\omega _n) \right ) ^{-1} \label{eq:greenscf}~~.
\end{eqnarray}
Eq.~(\ref{eq:greenscf}) is shown diagrammatically  in Fig. \ref{fig:G4SCC},
and from now on we refer to this level of approximation as SCC. (Note that 
we find that nothing new is learned when one examines the self-consistent,
non conserving level of approximation, and in this paper we ignore
such equations.) The significant achievement of a self-consistent
calculation is the inclusion of pair-pair interactions, as was
stated previously by Haussmann \cite{haussmann93}; unlike Haussmann,
who studied a 3D continuous system, here we study a 2D lattice system,
and are thus able to critique the physics of Ref. \cite {svr}.

Equations (\ref{eq:chi},\ref{eq:gamma},\ref{eq:sigscf},\ref{eq:greenscf}) 
have to be solved in an iterative way until self consistency is
achieved. Our procedure for accomplishing this, along the real time
axis, is discussed in the next section.

\section{Calculation procedure}
\label{sec:calcs}
\subsection{Formalism:}
\label{subsec:formalism}

In this section we outline how we solved the self-consistency problem
in our ${\bf k}$-averaged approximation. Also, we have employed 
another approximation in the spectral representations for certain
functions, and here we make this approach clear. Lastly, as was mentioned
earlier, our work on this problem involves the real-frequency 
formulation of the thermal Green's functions. This formulation
follows naturally from the analyticity of the retarded (and advanced)
Green's function which is, {\it e.g.}, explained in the work 
of Zubarev \cite{zubarev60},
and we refer the reader to this reference for further details.
The work in this section makes clear the relationship between
this formalism and the Matsubara frequency formalism.

In order to solve for the non self-consistent version of the ladder
approximation, in either conserving or non-conserving theories,
one uses the lattice Green's function to evaluate
the pair susceptibility. The susceptibility is used to calculate the
vertex, from which 
one can evaluate the self energy. In order to solve for these equations
self consistently, one uses the self energy to evaluate a new
approximation for the Green's function, and repeats the above process until
the resulting Green's function converges. 

In our ${\bf k}$-averaged approximation, we require only
the momentum averaged pair susceptibility; thus, only the
momentum averaged Green's function is required. That is, denoting the 
${\bf k}$-averaged quantities by overlined quantities
\begin{eqnarray}
\label{eq:chikav} 
\overline{\chi}(i \Omega _n)&\equiv&\frac{1}{N}\sum_{\bf K}\chi
({\bf K},i \Omega _n) \\
&=& -{1\over T} \sum_{m} 
\overline{G} (i\Omega_n - i\omega_m) \overline{G} (i\omega_m) 
\end{eqnarray}

As a computational approximation, we use a spectral representation for the 
${\bf k}$-averaged functions: they are
approximated by a number of $\delta$-functions (typically several
hundred) along the real axis. These $\delta$-functions were placed in 
a fashion such that they are exponentially dense around the chemical 
potential (with a spacing $\Delta$E $\ll k_B T$) and around the lower
band edge. The advantage of this representation 
is that all frequency summations appearing in our ${\bf k}$-averaged 
approximation can be done analytically.

According to our spectral representation, the averaged one-particle Green's 
function reads:
\begin{equation}
\label{gdn}
\overline{G}^{\tilde{m}}(i \omega _n) = 
\frac{1}{\pi} \int \frac{A(\omega)}{i \omega _n - \omega} d \omega 
\approx \sum_j^{N(\tilde{m})} \frac{a_j^{\tilde{m}}}{i \omega _n - b_j^{\tilde{m}}} 
\end{equation}
To express the Green's function entirely by a series of poles along the
real axis where to every frequency can belong a superposition of different
degenerate energy levels, is usually called the Lehmann representation
\cite{fetwal297ff,lehmann54}. In this way we also understand our
approximation above.

The function $A(\omega)$ is the imaginary part of the (retarded) 
one-particle Green's function along the real axis \cite{zubarev60,tyablikov67}. 
The superscript $(\tilde{m})$ in this and future quantities
labels the number of the iteration step
as convergence to self consistency is performed. Further, $N(\tilde{m})$ 
is the number of $\delta$ functions which were used, $b_j^{\tilde{m}}$ is 
the position of each $\delta$ peak on the real axis, and $a_j^{\tilde{m}}$ 
is the weight of this peak.

If we define the averaged Green's function in this way, we can
calculate analytically the frequency summation which is needed to
obtain the averaged pair susceptibility 
$\overline{\chi}^{\tilde{m}}(i \Omega _n)$:  
\begin{eqnarray}
\label{chispr}
\lefteqn{
\overline{
\chi}^{\tilde{m}}(i \Omega _n) = - \sum_{j,k}^{N(\tilde{m})} \frac{1}{\beta} \sum_m 
\frac{a_j^{\tilde{m}}}{i \omega _m - b_j^{\tilde{m}}} \frac{a_k^{\tilde{m}}}{i \Omega _n - i \omega
_m - b_k^{\tilde{m}}}} \nonumber \\
&=&
\sum_{j,k}^{N(\tilde{m})}  
\frac{a_j^{\tilde{m}} \; a_k^{\tilde{m}} }{i \Omega _n - b_j^{\tilde{m}} - b_k^{\tilde{m}}} \left (
\frac{1}{1+e^{- \beta b_j^{\tilde{m}}}} - \frac{1}{1+e^{- \beta b_k^{\tilde{m}}}}
\right )  \nonumber \\
&=&
\frac{1}{2} \sum_{j,k}^{N(\tilde{m})} \frac{a_j^{\tilde{m}} \; a_k^{\tilde{m}} }{i \Omega _n - b_j^{\tilde{m}} -
b_k^{\tilde{m}}}
\nonumber \\ && \;\;\;\;
 \left ( \tanh \left ( \frac{\beta b_j^{\tilde{m}}}{2} 
\right ) + \tanh \left ( \frac{\beta b_k^{\tilde{m}}}{2} 
\right ) \right )  
\end{eqnarray} 
which can be abbreviated as:
\begin{equation}
\label{eq:chisp}
\overline{\chi}^{\tilde{m}}(i \Omega _n) = \sum_k^{M(\tilde{m})} 
\frac{c_k^{\tilde{m}}}{i \Omega _n - 
d_k^{\tilde{m}}}
\end{equation} 
{\em viz.}, is of an identical formal structure as the ${\bf k}$-averaged
Green's function in our spectral representation but the spectral
weights of the poles are now given by $c_k^{\tilde{m}}$ at positions 
$d_k^{\tilde{m}}$ along the real axis.
Here $M(\tilde{m}) = \frac{N(\tilde{m})(N(\tilde{m})+1)}{2}$ is the 
new number of poles that are included in the representation for the
pair susceptibility. 

With $\overline{\chi}^{\tilde{m}}(i \Omega _n)$ we can now calculate
$\overline{\Gamma}^{\tilde{m}}(i \Omega _n)$, which in our theory is
given by
\begin{eqnarray}
\overline{\Gamma}(i \Omega _n)&=&\frac{1}{N}\sum_{\bf K}\Gamma
({\bf K},i \Omega _n) \nonumber \\ & \approx & \frac{U^2 \overline{\chi}
(i\Omega_n)  }{(1 + \mid U \mid \overline{\chi}
(i\Omega_n) ) } \nonumber \\ && \;\;
~[ 1 ~+~ {U ({\overline \chi^2} - {\overline {\chi^2}}) 
\over {\overline \chi} (1 + \mid U \mid {\overline \chi})} + \; ... ]~~.
\label{eq:gammakav}
\end{eqnarray}
We assume that the second term in the expansion, proportional to
the mean-squared fluctuations of the pair susceptibility,
and higher order terms, can be neglected based on our knowledge that they
should tend to zero in infinite spatial dimensions.
In order to obtain a spectral representation for
$\Gamma$ we use partial fractions:
\begin{eqnarray}
\label{eq:gamsp}
\lefteqn{
\overline{\Gamma}^{\tilde{m}}(i \Omega _n) = \frac{U^2 \overline{\chi}^{\tilde{m}}(i \Omega _n)}
{1- U \; \overline{\chi}^{\tilde{m}}(i \Omega _n) }} \nonumber \\
&=& \frac{U^2 \sum_k^{M(\tilde{m})} c_k^{\tilde{m}} \prod_{l \neq k} (i \Omega_m - d_l^{\tilde{m}})}
{\prod_k (i \Omega_m - d_k^{\tilde{m}}) - U \sum_k^{M(\tilde{m})} c_k^{\tilde{m}} \prod_{l \neq k} (i
\Omega_m - d_l^{\tilde{m}})} \nonumber \\
&=& \sum_m^{M(\tilde{m})} \frac{g_m^{\tilde{m}}}{i \Omega _n - h_m^{\tilde{m}}}
\end{eqnarray} 
The position of the poles $h_m^{\tilde{m}}$ along the real axis is
given by the zeroes of the polynomial 
\begin{equation}
\label{eq:gampfrac}
\prod_k (x - d_k^{\tilde{m}}) - U \sum_k^{M(\tilde{m})} c_k^{\tilde{m}} \prod_{l \neq k} (x
- d_l^{\tilde{m}}) = 0
\end{equation} 
which have to be determined numerically. The weight factors $g_m^{\tilde{m}}$
follow from the equation:
\begin{equation}
\label{eq:gamres}
g_m^{\tilde{m}} = U^2 \sum_k^{M(\tilde{m})} c_k^{\tilde{m}} \prod_{l \neq k} (h_m^{\tilde{m}} - d_l^{\tilde{m}})
\end{equation} 

With the result for $\Gamma$ we obtain an equation for the spectral 
representation of $\Sigma$. 
\begin{eqnarray}
\label{eq:sigspr}
\lefteqn{\overline{\Sigma}^{\tilde{m}}(i \omega_n) = \sum_r^{M(\tilde{m})} \sum_s^{N(\tilde{m})}
\frac{1}{\beta} \sum_m 
\frac{g_r^{\tilde{m}}}{i \omega _n + i \omega_m - h_r^{\tilde{m}}} 
\frac{a_s^{\tilde{m}}}{i \omega _m - b_s^{\tilde{m}}}} \nonumber \\
&=& \sum_r^{M(\tilde{m})} \sum_s^{N(\tilde{m})} 
\frac{g_r^{\tilde{m}} a_s^{\tilde{m}}}{i \omega _n - h_r^{\tilde{m}} + b_s^{\tilde{m}}} \left (
\frac{1}{1+e^{\beta b_s^{\tilde{m}}}} - \frac{1}{1-e^{\beta h_r^{\tilde{m}}}} \right )
\end{eqnarray} 
Again we can abbreviate this as
\begin{equation}
\label{eq:sigres}
\overline{\Sigma}^{\tilde{m}}(i \omega_n) = \sum_t^{N(\tilde{m}) \; M(\tilde{m})}
\frac{s_t^{\tilde{m}}}{i \omega _n - t_t^{\tilde{m}}}
\end{equation} 

To determine the equation for G in the next level of iteration we 
use partial fractions again:
\begin{eqnarray}
\label{greennp1}
\lefteqn{\overline{G}^{n+1}(i \omega_n) = \sum_j^{N(0)} 
\frac{a_j^0}{i \omega _n - b_j^0 - 
\overline{\Sigma}^{\tilde{m}}(i \omega_n)
}} \nonumber \\
&=& \sum_i^{N(0)} \frac{a_i^0 \prod_t (i \omega _n - t_t^{\tilde{m}} )}{
C(i \omega _n) - D(i \omega _n)} 
\nonumber \\ 
&=& \sum_i^{N(\tilde{m}+1)} \frac{a_i^{n+1}}{i \omega _n - b_i^{n+1}}      
\end{eqnarray}
Where we used the abbreviations:
\begin{eqnarray}
C(i \omega _n) &=& (i \omega _n - b_j^0) \prod_t (i \omega _n -
t_t^{\tilde{m}} )
\nonumber \\
D(i \omega _n) &=& 
\sum_t^{N(\tilde{m}) \; M(\tilde{m})} s_t^{\tilde{m}} \prod_{u \neq t}
(i \omega _n - t_u^{\tilde{m}} ) 
\end{eqnarray} 
The poles $b_j^{n+1}$ for the spectral representation of
$\overline{G}^{n+1}(i \omega_n)$ are given by the zeroes of the
polynomial
\begin{equation}
\label{greenspr}
(i \omega _n - b_j^0) \prod_t (i \omega _n - t_t^{\tilde{m}} ) -
\sum_t^{N(\tilde{m}) \; M(\tilde{m})} s_t^{\tilde{m}} \prod_{u \neq t}
(i \omega _n - t_u^{\tilde{m}} ) 
= 0
\end{equation} 
and the weight factors $a_j^{n+1}$ are obtained by inserting the
results for the poles in Eq. (\ref{greennp1}). 

By going through one loop of
self-consistency in this equations the number of poles is increased
from $N(\tilde{m})$ to 
$N(0) N(\tilde{m}) \; M(\tilde{m}) = N(0) N(\tilde{m})^2(N(\tilde{m}) +1) / 2$. 
To avoid that the number of
$\delta$-functions we have to deal with exceeds the number we can
handle numerically and to avoid divergences which can occur if two
poles come too close to each other we apply two additional
approximations.
First, we unite two delta peaks to a single one if their positions come
closer to each other then $\epsilon(\omega)$, where $\epsilon$ is smallest at
the chemical potential and at the lower band edge of the unperturbed
system. Second, we neglect a pole whose weight is smaller than a certain
boundary $\nu(\omega)$ which is again smallest at the chemical
potential. This has to be done in the way that the loss of spectral
weight is distributed onto all other poles in order to fulfill
the sum rules.

In this way we have solved Eq. (\ref{chispr}) to Eq. (\ref{greenspr}) 
until a stable self-consistent solution is obtained. $N(0)$ is
typically chosen to be around 20, and we end up with $N(\tilde{m}_{max})$ 
of the order of 300 and at no step of the calculation does
the number of poles exceeds 3000. 

We can therefore do the whole self-consistency loop by calculating all
quantities {\bf {\it along the real axis}}. In order to show that this is
indeed equivalent to the imaginary axis Matsubara frequency formalism
we show how one can calculate the particle number in both formalisms.
The expectation value of the particle number, $<n>$,
can be calculated either by summing over the poles along the imaginary
axis or by summing over the $\delta$ functions along the real axis:
\begin{equation}
<n> = \frac{1}{\beta} \sum_{\ell=-\infty}^{\infty} G(i\omega_\ell) 
= 1 + \frac{2}{\beta} \sum_{\ell=0}^{\infty} \Re e(G(i\omega_\ell))
\end{equation}
which can be expressed by using the spectral representation as an
integration along the real axis: 
\begin{eqnarray}
<n> &=& \lim_{\delta \rightarrow 0^+} \frac{1}{2  \pi i}
\oint \frac{e^{i \delta}}{1 + e^{\beta z}} G(z) \; dz \nonumber \\
&=& \lim_{\delta \rightarrow 0^+} \frac{1}{2 \pi i}
\oint \frac{e^{i \delta}}{1 + e^{\beta z}} 
\frac{1}{\pi} \int_{-\infty}^{\infty} 
\frac{A(x)}{z-x} \; dx \; dz \nonumber \\
&=& \frac{1}{\pi} \int_{-\infty}^{\infty} 
\frac{A(x)}{1 + e^{\beta x}} \; dx
\end{eqnarray}
This can be reduced by approximating the spectral function as a sum of
$\delta$ functions as done above:
\begin{equation} 
A(x) \approx \sum_j^{N(\tilde{m})} a_j^{\tilde{m}} \; \pi \;
\delta(x-b_j^{\tilde{m}})
\end{equation}
Together we can express the particle number with these
$\delta$ functions:
\begin{eqnarray}
<n> &\approx& \frac{1}{\pi} \int_{-\infty}^{\infty} 
\frac{1}{1 + e^{\beta x}}
\sum_j^{N(\tilde{m})} a_j^{\tilde{m}} \; \pi \;
\delta(x-b_j^{\tilde{m}}) \; dx \nonumber \\
&=& \sum_j^{N(\tilde{m})} 
\frac{a_j^{\tilde{m}}}{1 + e^{\beta \; b_j^{\tilde{m}} }} 
\end{eqnarray}

Note that the dimensionality of the system enters into this
calculation as the shape of $G^0$, whose imaginary part is the 
spectral function of the uncorrelated system. To obtain numerical
results we composed the unperturbed Green function $G^0$ of a 
number (typically 20) of $\delta$-functions. 
Since the most important physics is happening at
the chemical potential and at the lower band edge we choose the
distance between the $\delta$-peaks to be smallest at these points. To make
sure that the distance between two $\delta$-peaks is always smaller
then $k_BT$, at these points we made these distances exponentially small
by sampling the points with an tanh$^{-1}(\beta*x)$-function around the 
chemical potential and around the lower band edge.

\subsection{Justification of the ${\bf k}$-averaged method:}
\label{subsec:justify}

In the following we discuss limiting cases where the ${\bf k}$-average
approximation made in Eq.~(\ref{eq:gammakav}) becomes manifestly justifiable.

This is the case for large temperatures since $1-\langle n _{{\bf
K}/2 + {\bf q}} \rangle -\langle n _{{\bf
K}/2 - {\bf q}} \rangle $, as the numerator for the calculation of
$\chi$, is small. Therefore $\chi$ becomes small. Of course, this
is simply the uncorrelated limit (note that a ${\bf k}$ dependence
still survives in the Green's function, since the free Green's
function has such a dependence).
The approximation also holds true trivially for small
bandwidth and becomes exact if the band can be approximated by a
$\delta$-function ({\it viz.}, the atomic limit, for
which $t \longrightarrow 0$). In this case no ${\bf k}$ dispersion 
is present in the problem. Further, the case $\mid U \mid$ = 0
is trivially correct.

\begin{figure}
\unitlength1cm
\epsfxsize=13cm
\begin{picture}(7,7.5)
\put(-4.2,-1){\rotate[r]{\epsffile{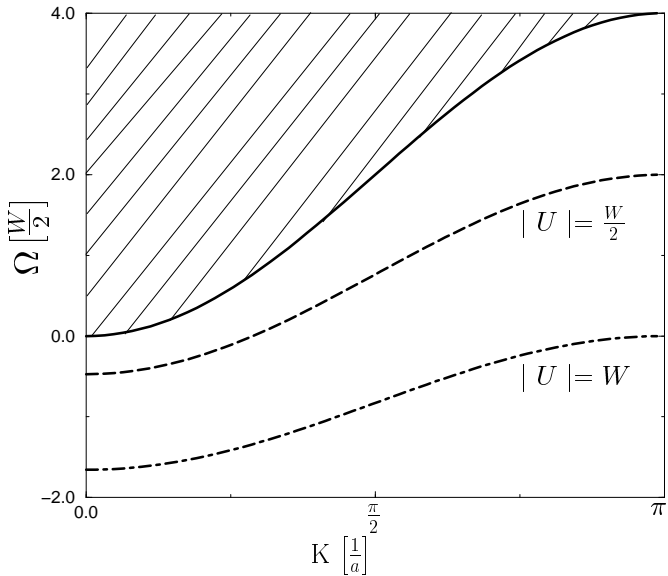}}}
\end{picture}
\caption{Schematic of the dispersion of the two-particle bound
state below the non-interacting continuum of the susceptibility 
$\chi(K,\Omega)$ (shaded region) as it arises in a NSC formulation, if
the chemical potential is below the continuum. The $k$-average approximation
is related to ignoring the dispersion of this band of states.}
\label{fig:2pdisp}
\end{figure}

For $\mid U \mid$ not too large, $\Gamma$ is determined by the pole of 
the bound state at $\chi = 1/\mid U \mid$, which means that a weakly dispersive 
two-particle bound state is well separated from the continuum. 
The dispersion becomes weaker for larger $\mid U \mid$ since the 
effective transfer of a pair is $t_{eff} \sim t^2 / U$.
So, in this case the average over ${\bf k}$-space is also a good 
approximation, since all we are doing is replacing the bound state
below the continuum by its average. This is shown schematically
in Fig.~\ref{fig:2pdisp}.

Probably the most interesting limit is that of large spatial
dimensions. It was argued by Metzner {\it et al.}  \cite{vollhardt90} that
for a system with large dimensions the ${\bf k}$ dispersion of the
self energy vanishes, which in the end is very similar to what
we do. Such an approximation gives several results which
are seemingly also valid in 2D and 3D; for a review see \cite{infdrev}. 

The only case where we can carefully critique the quality of our
fully self-consistent results found using our $k$-averaged
method is for a 1D model system. Of course, 
as we argued in relation to Eq.~(\ref{eq:gammakav}), in this dimension our
results should be least accurate due to large fluctuation effects.
Therefore, such a comparison should be viewed as an upper bound
of the potential differences, and our method should work much better
in any higher dimension. 

In the 1D case we can directly compare results obtained by our $k$-averaged 
method (we started with a density of states 
$A^0(\omega) = 1 / (\pi \sqrt{4t^2 - \omega ^2 })$ 
and 20 initial $\delta$-functions)
with fully self-consistent calculations (obtained by summing over the
Matsubara frequencies --- we used 40 to 60 lattice points and 150 fermionic and
299 bosonic Matsubara frequencies). Of course, for the latter
calculation, in order to perform this comparison we must restrict
ourselves to and calculate quantities that are averaged over the Brillouin 
zone.

\begin{figure}
\unitlength1cm
\epsfxsize=13cm
\begin{picture}(7,7.5)
\put(-4.2,-1){\rotate[r]{\epsffile{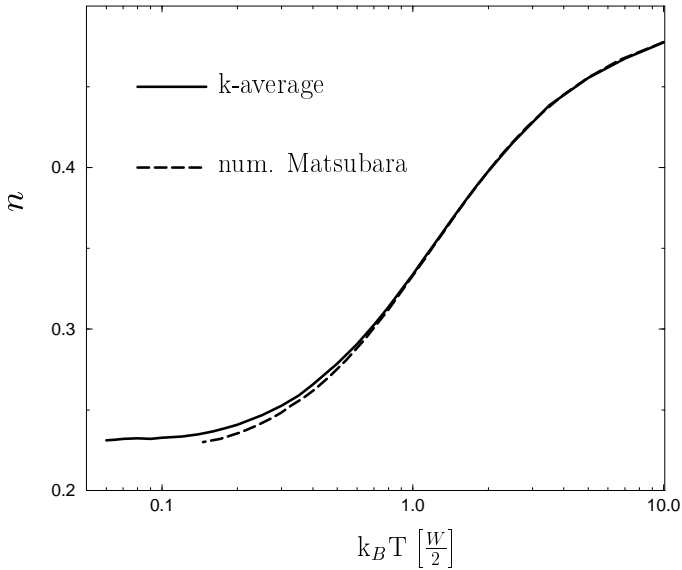}}}
\end{picture}
\caption{For a fixed chemical potential ($\mu$ = -0.9) the particle number
is plotted as a function of the temperature. All quantities are measured in
units of half the bandwidth, $W = 4t$ for 1D, and a value of $\mid U \mid = W$
is used. The k-average method is compared with
a numerical summation on a finite lattice (60 sites) and a finite
number of Matsubara frequencies (150). Both methods give essentially
the same result. Note the logarithmic temperature scale.} 
\label{fig:nvsT1d}
\end{figure}

In Fig. \ref{fig:nvsT1d} we have plotted the particle number as a
function of the temperature for a fixed chemical potential
obtained from the fully self consistent calculations. This was done for the
two entirely different calculational procedures. As one can see in
this figure, we had
to use a logarithmic scale to show the differences between these
two results. One curve is obtained by applying the $k$-averaged method
while the other curve is obtained taking into account the full
$k$-dispersion and summing numerically over the Matsubara frequencies.
Both methods describe the same physics, which is different from the
results obtained from NSC calculations.
That these curves agree so well, and the fact that we are able
to use our $k$-averaged method to such low temperatures, is very encouraging.

\begin{figure}
\unitlength1cm
\epsfxsize=11cm
\begin{picture}(7,7.5)
\put(-2.7,0){\rotate[r]{\epsffile{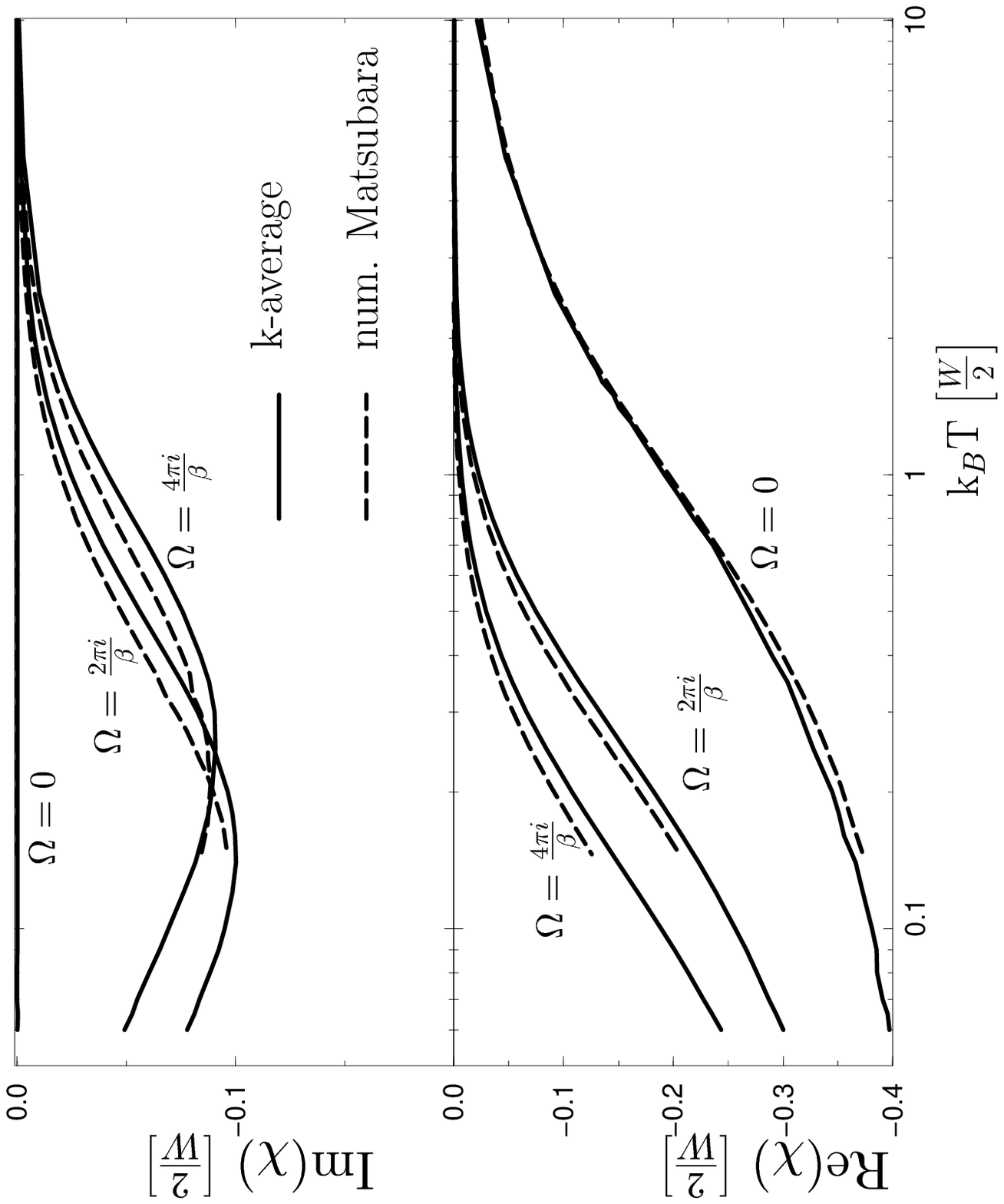}}}
\end{picture}
\caption{For three Matsubara frequencies ($\{0,\frac{2
\pi}{\beta},\frac{4 \pi}{\beta}\}$) and for the same 1D system the
susceptibility $\overline{\chi}$ is compared as a function of temperature. 
The two different approximations, full k-dependence numerical sum over
Matsubara frequencies and k-averaged method are compared. In the upper
graph the imaginary part of $\overline{\chi}$ is shown and in the
lower graph the real part. }
\label{fig:chivsT1d}
\end{figure}

We compared $k$-averaged susceptibilities for three
Matsubara frequencies, $ \Omega_n \in \{0,\frac{2 \pi i}{\beta},
\frac{4 \pi i}{\beta} \}$. The results in Fig. \ref{fig:chivsT1d} 
are obtained with the $k$-averaged method ---
note that the analyticity of $\chi(z)$ allows us to evaluate also the
results for the $k$-averaged method for imaginary frequencies.

The curves in Fig. \ref{fig:chivsT1d} are obtained by calculating all 
quantities by summing over (intermediate) Matsubara frequencies and 
by considering the full $k$ dispersion during the entire self-consistency 
calculation; then, the $k$-averaged is done at the very end of the 
calculation, after self-consistency had been established. The temperature 
dependence of these susceptibility quantities shows an excellent
agreement for these two completely different calculation methods.
Thus, the validity of both the $k$-average method, and of the
approximation discussed in Eq.~(\ref{eq:gammakav}), seems to be fine,
and our method successfully reproduces the full self-consistent
calculation even in 1D, where the fluctuations in Eq.~(\ref{eq:gammakav})
are expected to be largest.

\begin{figure}
\unitlength1cm
\epsfxsize=10cm
\begin{picture}(7,8.5)
\put(-2.7,0){\rotate[r]{\epsffile{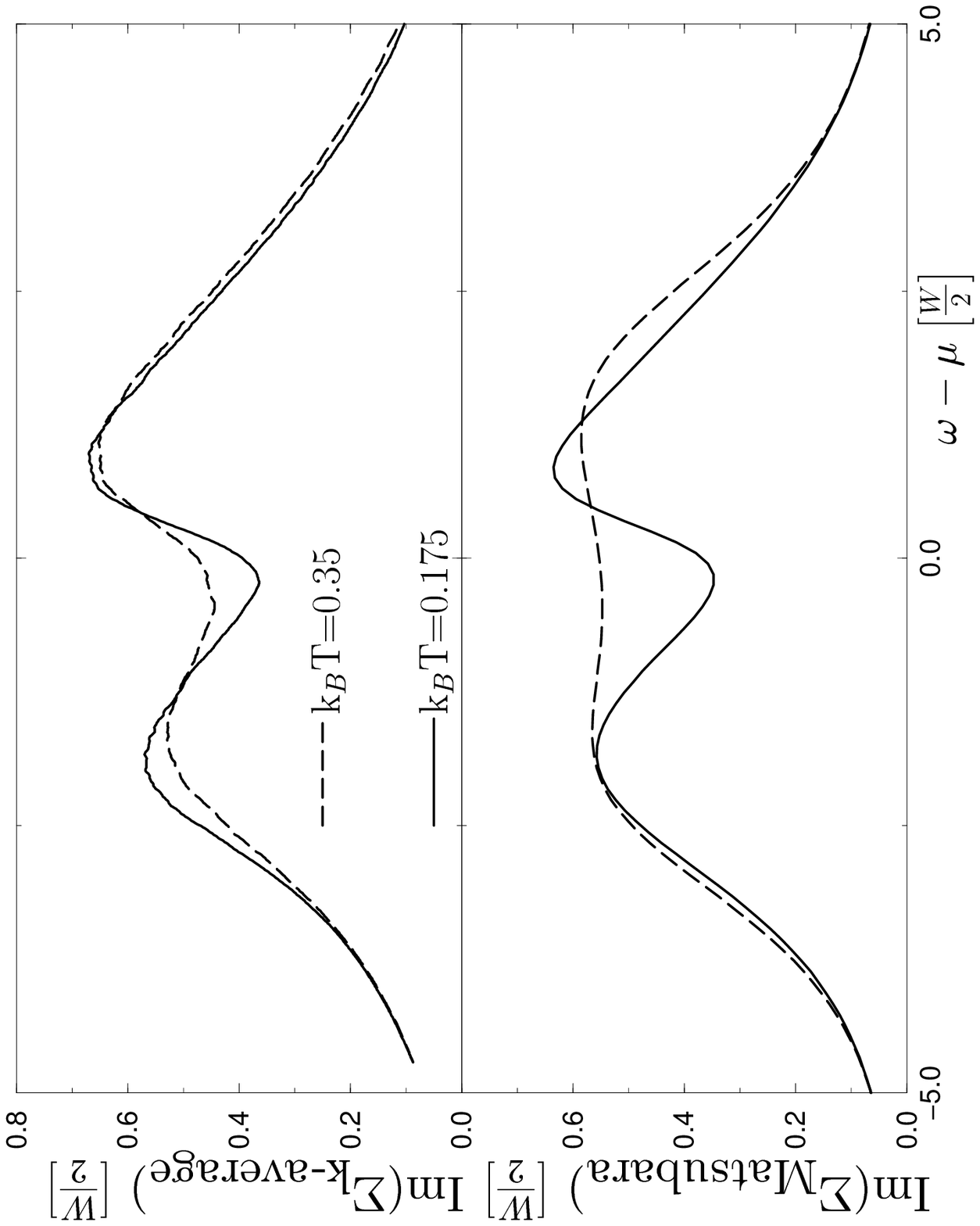}}}
\end{picture}
\caption{The k-averaged imaginary part of $Im(\Sigma(\omega-\mu))$ 
for a 1D system is
plotted. In the upper graph the k-averaged method is used whereas in
the lower graph a numerical sum over Matsubara frequencies (150
points) and 60 k-points with a k-average applied at the very end of
the calculation. In the latter case Pade approximates were used to
obtain real axis results. The difference at $k_BT = 0.35$ is due to the
failure of the Pade approximate at sparse Matsubara frequencies. The
minimum of $Im(\Sigma(\omega-\mu))$ is clearly resolved in both cases
for low temperatures. 
}
\label{fig:self1dcmp}
\end{figure}

To critique the predictions of dynamical quantities produced by the
$k$-averaged method, 
in Fig. \ref{fig:self1dcmp} we compare, for two different
temperatures, the imaginary parts of the $k$-averaged self energy
obtained in Fig. \ref{fig:self1dcmp}a by using the $k$-averaged
method, and in Fig. \ref{fig:self1dcmp}b by applying Pad\`e approximants,
as explained in Ref. \cite{vidberg_serene}, for
Matsubara frequencies where the full $k$-dependence has been
considered. 

\begin{figure}
\unitlength1cm
\epsfxsize=10cm
\begin{picture}(7,8.5)
\put(-2.2,0){\rotate[r]{\epsffile{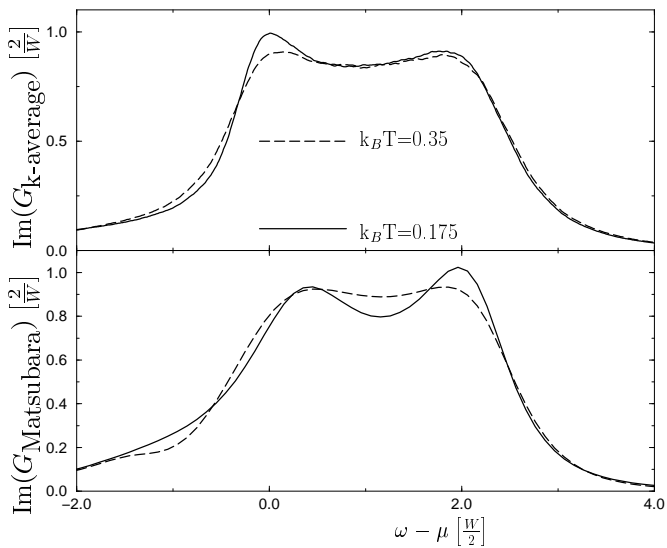}}}
\end{picture}
\caption{The 1D density of states for the two different approximations is
plotted for the same temperatures then
Fig. \protect\ref{fig:self1dcmp}. Note the 
local maximum at the chemical potential.}
\label{fig:dos1dcmp}
\end{figure}

In a similar fashion we compare in Fig. \ref{fig:dos1dcmp} 
the results for the one particle density of states. Again, the $k$ averaging
for the Matsubara frequency method was been done at the end of
the full self-consistent calculation, while the $k$-averaged method uses 
$k$-averaged quantities only throughout its approach to self consistency. 
In both of these figures we find good agreement between the two methods,
showing that the more conventional Matsubara frequency method results
are reproduced by our $k$-averaged approach to the pair susceptibility.

Encouraged by these successes of our $k$-averaged method in 1D, 
below we consider the system of greater interest, two dimensions.

\section{Results}
\label{sec:results}

In this section we discuss our numerical results obtained
using the method described above. We focus on a quasi 2D system 
by starting the $k$-averaged calculation with a $\overline{G^0}(\omega)$
that corresponds to a constant density of states. The band fillings are
chosen to be below half filling ($n$=0.5 is half filling), 
and therefore the absence of the van-Hove singularity in the middle of 
our ``band" is unimportant. We choose constant particle
numbers ($n$=0.1 and $n$=0.3) and calculate the chemical potentials as a
function of the temperature. 

\subsection{Single-particle density of states --- no pseudo gap}
\label{subsec:res4}

Of prime interest in our study is the single-particle density
of states, since the presence of a pseudo gap should be apparent
in this quantity \cite{swave}. 
Fortunately, this is a $k$-averaged quantity, and so it follows 
immediately from the imaginary part of the self-consistent
($k$-averaged) single-particle Green's function. 

\begin{figure}
\unitlength1cm
\epsfxsize=13cm
\begin{picture}(7,7.5)
\put(-4.2,-1){\rotate[r]{\epsffile{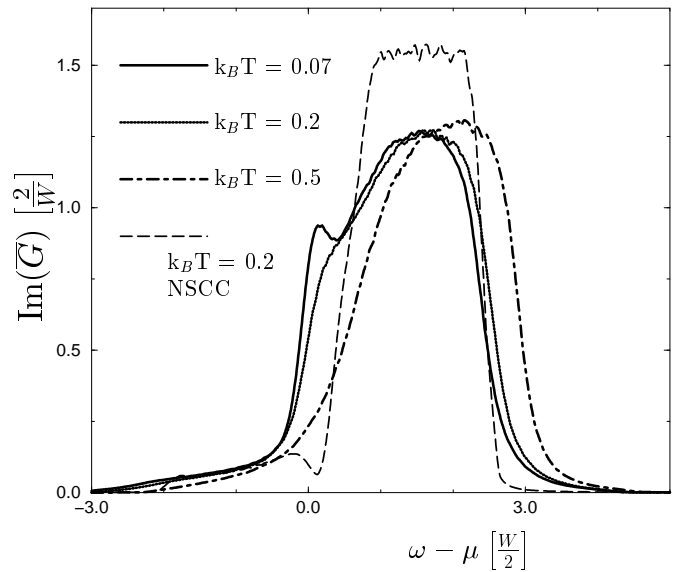}}}
\end{picture}
\caption{The resulting self-consistent density of states for the quasi two
dimensional system with $n$=0.1 and $\mid U \mid =2$ for four different 
temperatures is shown. Note that the density of states develops a maximum 
at the chemical potential when the temperature is decreased. For comparison
we have shown the density of states
from the NSCC approximation for $k_BT$ = 0.2.
All energies are given in units of $\left [ \frac{W}{2} \right ]$.} 
\label{fig:dos2dnpt1}
\end{figure}

\begin{figure}
\unitlength1cm
\epsfxsize=13cm
\begin{picture}(7,8.5)
\put(-4.2,-1){\rotate[r]{\epsffile{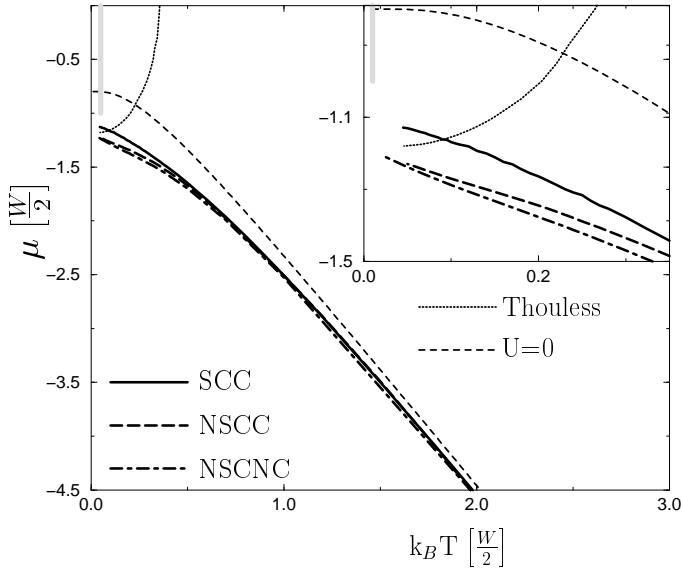}}}
\end{picture}
\caption{Flow diagram of the quasi 2D system ($n$=0.1) for the
chemical potential 
$\mu$ as a function of the temperature. The different levels of
approximation are shown.
For comparison, the line for the non-interacting
system (U=0) is given, and the Thouless line which determines the NSC
T$_c$ is shown. The region of the one particle
continuum of a non-interacting reference system is shown,
beginning at $\mu = -1$ on the $\mu$ axis.
The parameters were chosen to be $\mid U \mid = 2$ and $n=0.1$ and all
energies are given in units of half the bandwidth 
$\left [ \frac{W}{2} \right ]$.
}
\label{fig:muvsT2dnpt1}
\end{figure}

Our results for $n=0.1$ as a function of temperature (and which implicitly 
include the temperature-dependent chemical potential) are shown in
Fig. \ref{fig:dos2dnpt1}. For comparison, we also show the NSCC result for 
one temperature. To aid in the understanding of these results, we show the 
behaviour of the chemical potential as a function of temperature 
in Fig. \ref{fig:muvsT2dnpt1} --- this
latter quantity will be discussed in considerable detail in the next
subsection of the paper.

The NSCC result (which is very similar to the NSCNC result) at $k_BT = 0.2$
shows that the entire single-particle band is above the chemical potential
(seen from roughly $k_BT = 0.5$ to $k_BT = 2.5$), and that below the chemical potential one
finds the effects of the two-particle bound state. We stress that the 
``gap" between these two features is entirely different from that found by 
Janko, {\it et~al.} (which they proposed to be a pseudo gap)
\cite {janko97}, in that they study a 3D system which, more importantly, has
the chemical potential inside the one-particle band. As is seen in our
Fig. \ref{fig:muvsT2dnpt1}, for our NSCC calculation at $n$=0.1 the chemical
potential is below the continuum arising from the single-particle states.
Thus, the gap feature seen in our figure near $\omega = \mu$ is not related 
to a pseudo gap. Instead, we believe that this behaviour is related
to the analytical work for the NSC theory described in Ref. \cite {kagan97}.

Most importantly, the fully SCC result shows no evidence of a pseudo gap. 
We, in fact, see the opposite of a pseudo gap, wherein there is an increase 
in spectral weight at the chemical potential. Unfortunately,
it is somewhat difficult to say what happens to the chemical potential for 
this density --- so, we now consider higher densities to clarify this 
situation.

\begin{figure}
\unitlength1cm
\epsfxsize=13cm
\begin{picture}(7,7.5)
\put(-4.2,-1){\rotate[r]{\epsffile{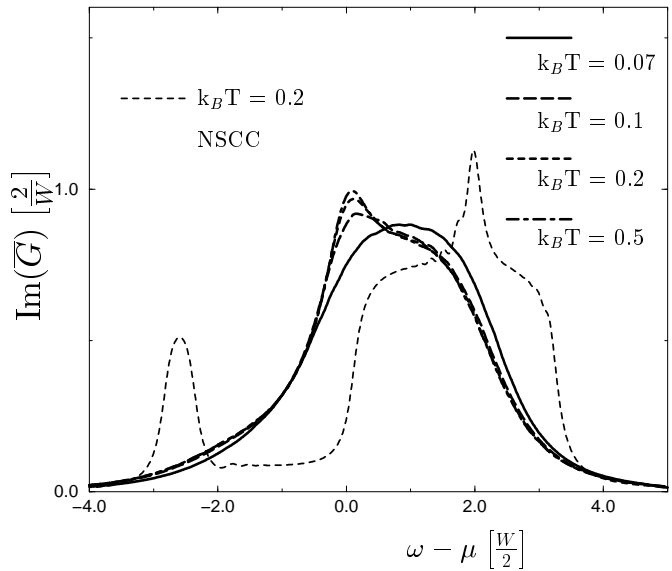}}}
\end{picture}
\caption{Same plot as Fig. \protect\ref{fig:dos2dnpt1} but for a density
of $n = 0.3$.}
\label{fig:dos2dnpt3}
\end{figure}

\begin{figure}
\unitlength1cm
\epsfxsize=13cm
\begin{picture}(7,7.5)
\put(-4.2,-1){\rotate[r]{\epsffile{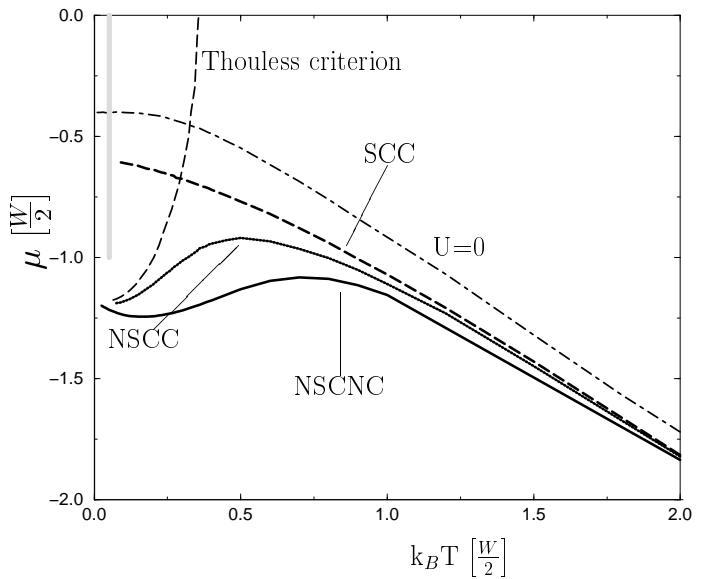}}}
\end{picture}
\caption{Same as Fig. \protect\ref{fig:muvsT2dnpt1} but for a density
of $n = 0.3$.}
\label{fig:muvsT2dnpt3}
\end{figure}

Figures \ref{fig:dos2dnpt3},\ref{fig:muvsT2dnpt3} show the analogous
results for $n$=0.3. Here, the NSCC result is similar to the above
NSCC data, only now the feature from the two-particle bound state
is much more clearly resolved. Also, the physics in the SCC result is 
more easily seen. These figures
show quite clearly that (i) the chemical potential is in the band
(for the SCC formulation, this is true for temperatures below about 0.9), 
and (ii) no evidence
of a pseudo gap is found. Instead, as above, there is an enhancement
of the spectral weight near the chemical potential. This finding is 
in disagreement with other SCC results 
for a $d$-wave pairing potential \cite {pseudonaz}.

\subsection{Suppression of superconductivity in 2D}
\label{subsec:res3}

As mentioned in the introduction, a simple explanation of the (potential)
suppression of superconductivity in 2D was given in Ref. \cite {svr}.
The physics can be related to the behaviour of chemical potential, namely,
when an attractive potential of any strength is present in 2D (or 1D)
a bound state is formed, and this state ``attracts" the chemical
potential to it. Thus, the chemical potential is found below the bottom
of the band, and no Fermi surface is left. No Fermi surface means
that there is no Cooper pairing. 

We note that the formulation of Ref. \cite{svr} is NSCNC. Although
Serene \cite {serene89} has claimed that the physics of Ref. \cite {svr}
is destroyed when one uses a conserving theory, we disagree with this.
In fact, for lattice systems, we have been able to formally show that
the physics of Ref. \cite {svr} survives, and in fact becomes more
transparent, in a conserving theory. We will discuss these results
in a future publication.

In contrast to this, our data for the chemical potential in a SCC theory
shows that the physics of Ref. \cite {svr} does not survive the
inclusion of pair-pair interactions in a fully self-consistent theory.
This is easiest to see in Fig. \ref{fig:muvsT2dnpt3}, wherein the
low temperature extrapolation of the chemical potential is into the
(interacting) 
single-particle band. Put another way, there is still a Fermi surface
at low temperatures in a SCC theory. (That is not necessarily to say
that this is a Fermi liquid --- see the next subsection.) We note
that the reappearance of the Fermi surface was also seen in the SCC 
work of Fresard et al. \cite {fresard92}.

\begin{figure}
\unitlength1cm
\epsfxsize=13cm
\begin{picture}(7,7.5)
\put(-4.2,-1){\rotate[r]{\epsffile{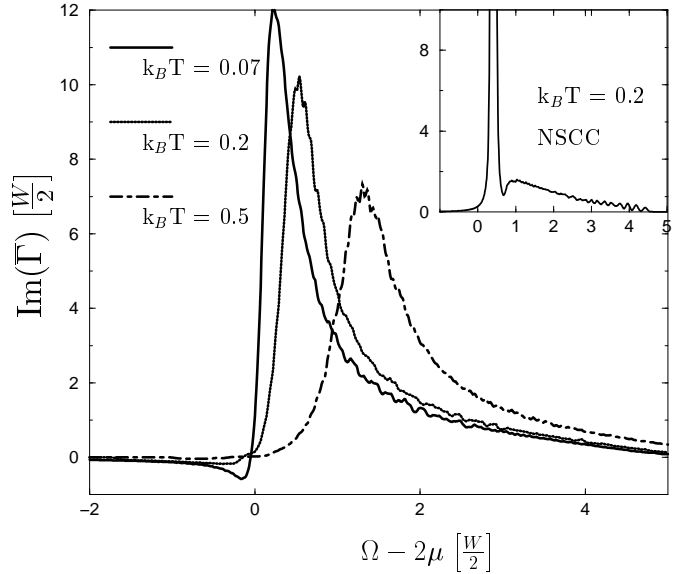}}}
\end{picture}
\caption{Plot of $Im(\overline{\Gamma}(\Omega-2\mu))$ for the quasi-2D
system. The inset shows
the result for $kT = 0.2$ from the NSCC calculation with an
delta peak indicating the infinite lifetime bound state. The SCC
result is plotted for three temperatures (0.5, 0.2, 0.07) and
shows no infinite lifetime bound state. All energies are again in
units of $\left [ \frac{W}{2} \right ]$ with $\mid U \mid = 2$ and $n$=0.1. 
}
\label{fig:imgam2dnpt1}
\end{figure}

To better understand this, we now consider the behaviour of this bound state 
via the SCC vertex function. Examining the vertex function 
$\overline{\Gamma} (\Omega)$ probably best explains the different physics 
between the NSCC and the SCC formulations:
For the NSCC calculation, the infinite lifetime bound state is seen
as a (numerically broadened) delta function in 
$Im(\overline{\Gamma} (\Omega))$ (inset in Fig. \ref{fig:imgam2dnpt1}).
However, $Im(\overline{\Gamma}(\Omega))$ for
the SCC calculation does not show a delta peak below the
continuum --- instead, at the chemical potential, with decreasing
temperature an enhancement near $\mu$ can be seen.
Note that in Fig. \ref{fig:imgam2dnpt1} for $\Omega$ below the chemical 
potential there is a non vanishing imaginary part of
$\overline{\Gamma}(\Omega)$ that is negative. Therefore, the
enhancement of $Im(\overline{\Gamma}(\Omega))$ can also be interpreted
as a lifetime broadened remnant of a two-particle bound state.
So, we stress again that for the SCC calculation there is
no infinite lifetime bound state, opposite to the NSCC calculation.
When treating the system self-consistently, pair-pair interactions
are included which lead to a finite lifetime of the pairs and therefore
the condensation of the fermions into a bound state is hindered.

\subsection{Quasi-particle lifetime at the chemical potential}
\label{subsec:qpaliti}

The above results showed that in a SCC theory the Fermi surface survives
at low temperatures.  The natural question is then: Is this a Fermi liquid?

\begin{figure}
\unitlength1cm
\epsfxsize=13cm
\begin{picture}(7,7.5)
\put(-4.2,-1){\rotate[r]{\epsffile{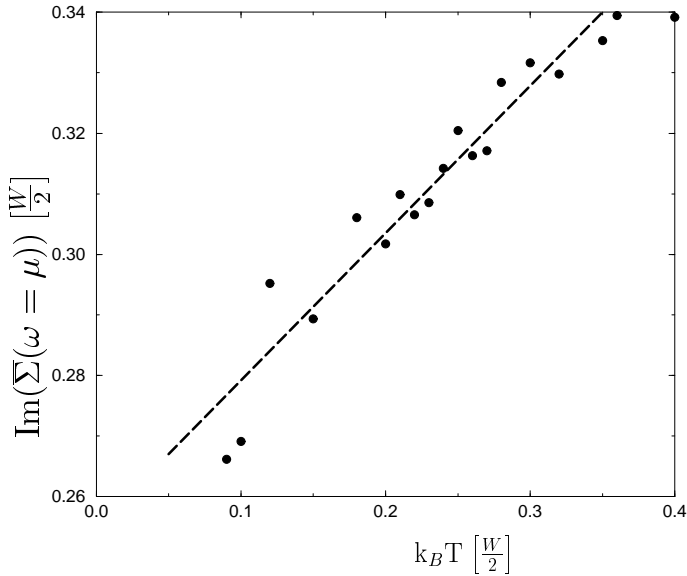}}}
\end{picture}
\caption{The imaginary part of the self energy at the
chemical potential as a function of temperature for
the model parameters $\mid U \mid = 2$ and $n$=0.1.
The straight line is a guide to the eye.}
\label{fig:sigmaimag2d}
\end{figure}

To help us answer this question,
we consider the (real) frequency dependence of the 
Imaginary part of the self energy $\Sigma$. In a SCC theory
it exhibits a minimum at the chemical potential, and this
qualitative behaviour is indeed similar to a Fermi liquid.
However, in Fig. \ref{fig:sigmaimag2d} we have plotted this quantity,
which is related to the inverse quasi-particle lifetime at the 
chemical potential, as a function of temperature. The
calculated value seems to indicate a linear variation with
temperature (the extrapolated value is simply a function of our 
numerical broadening, and in this calculation has no physical significance).

To understand this we note that standard
arguments, based on, e.g.,  Fermi's golden rule, predict the $T^2$ behaviour
of a Fermi liquid. However, and we argue, that if the dominant scattering is 
between lifetime broadened two-particle bound states and quasiparticles,
then such a linear T behaviour is indeed expected. Crucial to this
argument is the approximate temperature independence of the lifetime
of the two-particle bound state, something that our numerics make clear.
Further consequences of this simple phenomenology are presently being 
explored.

\section{Conclusion}
\label{sec:conclusion}

We have examined the attractive Hubbard model in 2D using a
${\bf k}$-averaged method. Further, we have studied
the dynamical properties of this model using real time axis
thermal Green's functions --- this allows for an accurate
determination of quantities like the single-particle density
of states, and the energy dependencies of the self energy
and vertex function. We have compared our approach to the
more familiar Matusbara frequency method in 1D, and have found
little difference between the results obtained by the two methods. 
However, the ${\bf k}$-averaged method allows us to reach
much lower temperatures without enormous computational efforts.

We have included pair-pair interactions, as formulated by Haussmann,
in a SCC theory. Our results for such a theory lead us to believe that 
the attractive Hubbard model in 2D, for a correlation energy roughly
equal to the bandwidth, does not have a pseudo gap. Instead, there is 
actually an enhancement of the spectral weight near the Fermi surface.
The nature of the ground state properties is uncertain, since we find a 
linear temperature dependence of the quasi-particle scattering rate over 
a wide temperature range. The reappearance of the Fermi surface in a SCC 
theory in 2D was noted previously by Fresard, et al. \cite {fresard92}, 
but the linear $T$ behaviour is new, and deserves further investigation.

\acknowledgements

We wish to thank R.~Haussmann, B.~Janko, M.~Kagan, S.~Kehrrein,
P.~Kornilovitch, A.~Nazarenko, M.~Randeria, M.~Ulmke, 
and especially F.~Marsiglio and A.~Chernyshev for helpful discussions.  
This work was supported by the ''Deutsche Forschungsgemeinschaft"
and by the NSERC of Canada.


\begin{thebibliography}{10}
\bibitem{swave}
Note that we are studying a system that would
display $s$-wave pairing, and thus one does not require the $k$
resolved density of states to see evidence of a pseudo gap.
\end{thebibliography}
\end{document}